\documentclass[12pt]{article}
\usepackage{layout,amsmath,amssymb}

\usepackage{epsfig,float}

\begin{document}

\title{\bf Matter Collineations of Static Spacetimes with
Maximal Symmetric Transverse Spaces}

\author{M. Sharif \thanks{msharif@math.pu.edu.pk}\\
Department of Mathematics, University of the Punjab,\\
Quaid-e-Azam Campus, Lahore-54590, Pakistan.}

\date{}

\maketitle

\begin{abstract}
This paper is devoted to study the symmetries of the energy-momentum
tensor for the static spacetimes with maximal symmetric transverse
spaces. We solve matter collineation equations for the four main
cases by taking one, two, three and four non-zero components of the
vector $\xi^a$. For one component non-zero, we obtain only one
matter collineation for the non-degenerate case and for two
components non-zero, the non-degenerate case yields maximum three
matter collineations. When we take three components non-zero, we
obtain three, four and five independent matter collineations for the
non-degenerate and for the degenerate cases respectively. This case
generalizes the degenerate case of the static spherically symmetric
spacetimes. The last case (when all the four components are
non-zero) provides the generalization of the non-degenerate case of
the static spherically symmetric spacetimes. This gives either four,
five, six, seven or ten independent matter collineations in which
four are the usual Killing vectors and rest are the proper matter
collineations. It is mentioned here that we obtain different
constraint equations which, on solving, may provide some new exact
solutions of the Einstein field equations.
\end{abstract}


\section{Introduction}

Let $M$ be a spacetime manifold with Lorentz metric $g$ of signature
(+ - - -). The manifold $M$ and the metric $g$ are assumed smooth
($C^{\infty}$). Throughout this article, the usual component
notation in local charts will often be used, and a covariant
derivative with respect to the symmetric connection $\Gamma$
associated with the metric $g$ will be denoted by a semicolon and a
partial derivative by a comma.

The Einstein's field equations (EFEs) in local coordinates are given
by
\begin{equation}
G_{ab} \equiv R_{ab} - \frac{1}{2} R g_{ab} = T_{ab},\quad
(a,b=0,1,2,3),
\end{equation}
where $G_{ab}$ are the components of the Einstein tensor, $R_{ab}$
those of the Ricci and $T_{ab}$ of the matter (energy-momentum)
tensor. Also, $R = g^{ab} R_{ab}$ is the Ricci scalar, and it is
assumed that $\kappa = 1$ and $\Lambda = 0$ for simplicity. In
General Relativity (GR) theory, the Einstein tensor $G_{ab}$ plays a
significant role, since it relates the geometry of spacetime to its
source.

The EFEs (1), whose fundamental constituent is the spacetime metric
$g_{ab}$, are highly non-linear partial differential equations, and
therefore it is very difficult to obtain their exact solutions.
Symmetries of the geometrical/physical relevant quantities of the GR
theory are known as {\it collineations}. In general, these can be
represented as $\pounds_{\xi} {\cal A} = {\cal B}$, where ${\cal A}$
and ${\cal B}$ are the geometric/physical objects, $\xi$ is the
vector field generating the symmetry, and $\pounds_{\xi}$ signifies
the Lie derivative operator along the vector field $\xi$.

A one-parameter group of conformal motions generated by a {\it
conformal Killing vector} (CKV) $\xi$ is defined as [1]
\begin{equation}
\pounds_{\xi} g_{ab} = 2 \psi g_{ab},
\end{equation}
where $\psi=\psi(x^a)$ is a conformal factor. If $\psi_{;\, a b}
\neq 0$, the CKV is said to be {\it proper}. Otherwise, $\xi$
reduces to the special {\it conformal Killing vector} (SCKV) if
$\psi_{;\,ab} = 0$, but $\psi_{,a}\neq 0$. Other subcases are {\it
homothetic vector} (HV) if $\psi_{,a}=0$ and {\it Killing vector}
(KV) if $\psi= 0$.

Using Eq.(2), we find from Eq.(1) that
\begin{equation}
\pounds_{\xi} T_{ab} = - 2 \psi_{;ab} + 2 g_{ab}\Box \psi  ,
\end{equation}
where $\Box$ is the Laplacian operator defined by $\Box \psi \equiv
g^{cd} \psi_{;cd}$. Therefore, for a KV, or HV, or SCKV we have
\begin{equation}
\pounds_{\xi} T_{ab} = 0 \quad \Leftrightarrow \quad \pounds_{\xi}
G_{ab} = 0,
\end{equation}
or in component form
\begin{equation}
\quad T_{ab,c} \xi^c + T_{ac} \xi^c_{,b} + T_{cb} \xi^c_{,a} = 0.
\end{equation}
A vector field $\xi$ satisfying Eq.(4) or (5) on $M$ is called a
{\it matter collineation} (MC). Since the Ricci tensor arises
naturally from the Riemann curvature tensor (with components
$R^a_{\, bcd}$ and where $R_{ab} \equiv R^c_{\, acb}$) and hence
from the connection, the study of {\it Ricci collineation} (RC)
defined by $\pounds_{\xi} R_{ab} = 0$ has a natural geometrical
significance [2]-[12]. Mathematical similarities between the Ricci
and energy-momentum tensors mean many techniques for their study
should show some similarity. Some papers have recently been appeared
on MCs [13]-[19]. From the physical viewpoint, a study of MCs, i.e.
to look into the set of solutions to Eq.(5) seems more relevant. In
addition, since energy-momentum tensor $T_{ab}$ is more fundamental
in the study of dynamics of fluid spacetimes of GR, the remainder of
this paper will be concerned with MCs.

The plan of the paper is as follows. In the next section we shall
write down MC Eqs.(5) for the maximal symmetric transverse metric.
In section 3, these equations are solved for different cases.
Finally, we shall provide a brief summary and discussion of the
results obtained.

\section{Matter Collineation Equations}

The general metric for a static spacetime with a maximal symmetric
transverse space is given by [20]-[22].
\begin{equation}
ds^2=e^{\nu(r)}dt^2-e^{\mu(r)}dr^2-r^2(d\theta^2+f^2(\theta)d\phi^2),
\end{equation}
where $f(\theta)$ is $\theta,~\sinh\theta$ or $\sin\theta$ according
as $k =-f_{,22}/f=0,~-1$ or $+1$ respectively. Notice that
$f^2(f_{,2}/f)_{,2}=-1$ if and only if $(ff_{,2})_{,2}=2f_{,2}^2-1$.
The non-zero components of the energy-momentum tensor for the above
metric are $ T_{00},~T_{11},~T_{22},~T_{33}$ given in Appendix $A$.
MC equations in component form can be written as
\begin{eqnarray}
T'_0\xi^1+2T_{0}\xi^0_{,0}=0,\\
T_{0}\xi^0_{,1}+T_{1}\xi^1_{,0}=0,\\
T_{0}\xi^0_{,2}+T_{2}\xi^2_{,0}=0,\\
T_{0}\xi^0_{,3}+f^2T_{2}\xi^3_{,0}=0,\\
T'_{1}\xi^1+2T_{1}\xi^1_{,1}=0,\\
T_{1}\xi^1_{,2}+T_{2}\xi^2_{,1}=0,\\
T_{1}\xi^1_{,3}+f^2T_{2}\xi^3_{,1}=0,\\
T'_{2}\xi^1+2T_{2}\xi^2_{,2}=0,\\
T_{2}(\xi^2_{,3}+f^2\xi^3_{,2})=0,\\
T'_{2}\xi^1+2\frac{f_{,2}}{f}T_2\xi^2+2T_2\xi^3_{,3}=0,
\end{eqnarray}
where prime denotes differentiation with respect to the radial
coordinate $r$. It is to be noted that we have used the notation
$T_{aa}=T_a$ for the sake of brevity.

\section{Solution of the MC Equations}

In this section, we solve the MC equations (7)-(16) for the
following four cases:\\
I.\quad One component of $\xi^a(x^b)$ is different from zero;\\
II.\quad Two components of $\xi^a(x^b)$ are different from zero; \\
III.\quad Three components of $\xi^a(x^b)$ are different from zero;\\
IV.\quad All components of $\xi^a(x^b)$ are different from zero.

\subsection{One Component of $\xi^a(x^b)$ is Different from
Zero}

This case has the following four possibilities: \\
(i)\quad $\xi^a=(\xi^0,~0,~0,~0)$;\\
(ii)\quad $\xi^a=(0,~\xi^1,~0,~0)$;\\
(iii)\quad $\xi^a=(0,~0,~\xi^2,~0)$;\\
(iv)\quad $\xi^a=(0,~0,~0,~\xi^3)$.\\
\par\noindent
In the case I(i), Eqs.(7)-(9) give $T_0\xi^0_{,a}=0$. It follows
that either $T_{0}=0$ or $T_{0}\neq0$. For $T_{0}=0$, we obtain
$\xi^0=\xi^0(x^a)$ and $T_{0}\neq0$ implies that $\xi^0$ is an
arbitrary constant.

For the case I(ii), Eqs.(9), (10) and (16) are identically satisfied
and the remaining equations become
\begin{eqnarray}
T'_0\xi^1&=&0=T'_{2}\xi^1,\\
T_{1}\xi^1_{,a}&=&0,\quad T'_{1}\xi^1+2T_{1}\xi^1_{,1}=0.
\end{eqnarray}
Eq.(17) implies that $T_0=constant=T_2$. Eq.(18) gives rise to
further four possibilities according to
\begin{eqnarray*}
&(a)&\quad T'_1=0,\quad T_{1}=0,\quad (b)\quad T'_{1}=0,\quad T_{1}\neq0,\\
&(c)&\quad T'_{1}\neq0,\quad T_{1}\neq0.
\end{eqnarray*}
For the first possibility, Eq.(18) implies that $\xi^1=\xi^1(x^a)$.
The second case Iii(b) gives $T_1=constant\neq0$ and hence
$\xi^1=constant$. In the third option Iii(c) when both
$T'_{1},~T_{1}\neq0$, it follows from Eq.(18) that
$\xi^1=\frac{c_0}{\sqrt{T_1}}$.

In the third case when $\xi^2\neq0$, it follows from MC equations
(9), (12), (14) and (15) that $T_2\xi^2_{,a}=0$. Also, Eq.(16)
yields $T_2\xi^2=0$ which implies that $T_2=0$ and hence
$\xi^2=\xi^2(x^a)$.

The last case I(iv), where $\xi^3\neq0$, MC equations (10), (13),
(15) and (16) yield $T_2\xi^3_{,a}=0$. This implies that either
$T_2=0$ or $T_2\neq0$. When $T_2=0$, we obtain $\xi^3=\xi^3(x^a)$
and for $T_2\neq0$, we get $\xi^3=constant$.

\subsection{Two Components of $\xi^a(x^b)$ are Different from Zero}

In this case, we have six different possibilities:\\
II(i)\quad $\xi^a=(\xi^0,~\xi^1,~0,~0)$;\\
II(ii)\quad $\xi^a=(\xi^0,~0,~\xi^2,~0)$;\\
II(iii)\quad $\xi^a=(\xi^0,~0,~0,~\xi^3)$;\\
II(iv)\quad $\xi^a=(0,~\xi^1,~\xi^2,~0)$;\\
II(v)\quad $\xi^a=(0,~\xi^1,~0,~\xi^3)$;\\
II(vi)\quad $\xi^a=(0,~0,~\xi^2,~\xi^3).$\\
\par\noindent
{\bf Case II(i):} $\xi^a=(\xi^0,~\xi^1,~0,~0)$\\
\par\noindent
In this case, Eq.(15) is identically satisfied. From Eq.(14) or
(16), it follows that $T'_2\xi^1=0$ which implies that
$T_2=constant=c~(say)$. From the remaining equations we have four
different options according to
\begin{eqnarray*}
&(a)&\quad T_0=0,\quad T_{1}=0,\quad (b)\quad T_{0}=0,\quad T_{1}\neq0,\\
&(c)&\quad T_{0}\neq0,\quad T_{1}=0,\quad(d)\quad T_{0}\neq0,\quad
T_{1}\neq0.
\end{eqnarray*}
For the first possibility, we obtain
$\xi^0=\xi^0(x^a),~\xi^1=\xi^1(x^a)$. In the second case, it follows
that $\xi^0=\xi^0(x^a),~\xi^1=\frac{c_0}{\sqrt{T_1}}$. For the third
option, we have $\xi^0=\xi^0(t),~\xi^1=\xi^1(x^a)$. Finally, in the
last possibility when both $T_0,~T_1\neq0$, we see from Eqs.(9),
(10) and (12), (13) that $\xi^0$ and $\xi^1$ become functions of
$t,r$ respectively and we are left with Eqs.(7), (8) and (11). From
Eq.(11) we have $\xi^1=\frac{A(t)}{\sqrt{|T_{1}|}}$, where $A(t)$ is
an integration function. Replacing this value of $\xi^0$ in Eq.(7),
it follows that
\begin{equation}
\xi^0(t,r)=-\frac{T'_{0}}{2T_{0}\sqrt{|T_{1}|}}\int A(t)dr+B(r),
\end{equation}
where $B(r)$ is an integration function. Substituting this value of
$\xi^0$ together with $\xi^1$ in Eq.(8), we obtain
\begin{equation}
\frac{\ddot{A}}{A}=\frac{T_{0}}{\sqrt{|T_{1}|}}
(\frac{T'_{0}}{2T_{0}\sqrt{|T_{1}|}})'=-\alpha^2,
\end{equation}
where $\alpha^2$ is a separation constant which may be positive,
negative or zero. \\
When $\alpha^2>0$, we have
\begin{eqnarray}
\xi^0&=&-\frac{T'_{0}}{2\alpha T_{0}\sqrt{|T_{1}|}}(c_0\sin\alpha
t-c_1\cos\alpha t)+c_2,\nonumber\\
\xi^1&=&\frac{1}{\sqrt{|T_{1}|}}(c_0\cos\alpha t+c_1\sin\alpha t),
\end{eqnarray}
where $c_0,~c_1,~c_2$ are arbitrary constants. It follows that MCs
can be written as
\begin{eqnarray}
\xi_{(1)}&=&\partial_t,\nonumber\\
\xi_{(2)}&=&\frac{T'_{0}}{2\alpha T_{0}\sqrt{|T_{1}|}}\sin\alpha
t\partial_t-\frac{1}{\sqrt{|T_{1}|}}\cos\alpha
t\partial_r,\nonumber\\
\xi_{(3)}&=&\frac{T'_{0}}{2\alpha T_{0}\sqrt{|T_{1}|}}\cos\alpha
t\partial_t+\frac{1}{\sqrt{|T_{1}|}}\sin\alpha t\partial_r.
\end{eqnarray}
When $\alpha^2<0$, $\alpha^2$ is replaced by $-\alpha^2$ in Eq.(21)
and we obtain the following solution
\begin{eqnarray}
\xi_{(1)}&=&\partial_t,\nonumber\\
\xi_{(2)}&=&\frac{T'_{0}}{2\alpha T_{0}\sqrt{|T_{1}|}}\sinh\alpha
t\partial_t-\frac{1}{\sqrt{|T_{1}|}}\cosh\alpha
t\partial_r,\nonumber\\
\xi_{(3)}&=&\frac{T'_{0}}{2\alpha T_{0}\sqrt{|T_{1}|}}\cosh\alpha
t\partial_t+\frac{1}{\sqrt{|T_{1}|}}\sinh\alpha t\partial_r.
\end{eqnarray}
For $\alpha^2=0$, we obtain
$\frac{T'_{0}}{2T_{0}\sqrt{|T_{1}|}}=\beta$, where $\beta$ is an
arbitrary constant. This implies that either $\beta$ is non-zero or
zero. For $\beta\neq 0$, it follows that
\begin{eqnarray}
\xi_{(1)}&=&\partial_t,\nonumber\\
\xi_{(2)}&=&(\beta\frac{t^2}{2}-\int\frac{\sqrt{|T_{1}|}}{T_{0}}dr)
\partial_t-\frac{t}{\sqrt{|T_{1}|}}
\partial_r,\nonumber\\
\xi_{(3)}&=&\beta t\partial_t-\frac{1}{\sqrt{|T_{1}|}}\partial_r.
\end{eqnarray}
For $\beta=0$, we have
\begin{eqnarray}
\xi_{(1)}&=&\partial_t,\nonumber\\
\xi_{(2)}&=&\frac{1}{T_{0}}\int{\sqrt{|T_{1}|}}dr\partial_t
+\frac{t}{\sqrt{|T_{1}|}}\partial_r,\nonumber\\
\xi_{(3)}&=&\frac{1}{\sqrt{|T_{1}|}}\partial_r.
\end{eqnarray}
Thus, in the subcase IIi(d), MCs turn out to be three in all the
possibilities. If Eq.(20) is not satisfied by $T_{0}$ and $T_{1}$
then $A=0$ and this reduces to the case I(ii). \\
\par\noindent
{\bf Case II(ii):} $\xi^a=(\xi^0,~0,~\xi^2,~0)$\\
\par\noindent
In this case, Eqs.(11) and (13) are identically satisfied and the
remaining equations reduce to
\begin{eqnarray}
T_{0}\xi^0_{,m}=0,\quad (m=0,1,3),\\
T_{0}\xi^0_{,2}+T_{2}\xi^2_{,0}=0,\\
T_{2}\xi^2_{,i}=0,\quad (i=1,2,3)\\
T_2\xi^2=0.
\end{eqnarray}
Eq.(29) implies that $T_2=0$ as $\xi^2\neq0$ and consequently, we
obtain $\xi^2$ as an arbitrary function of four-vector. Eqs.(26) and
(27) give rise to two possibilities either $T_0=0$ or $T_0\neq0$.
For the first possibility, $\xi^0$ becomes arbitrary function and
for the second case, $\xi^0$ turns out to be
arbitrary constant.\\
\par\noindent
{\bf Case II(iii):} $\xi^a=(\xi^0,~0,~0,~\xi^3)$\\
\par\noindent
In this case, Eqs.(11), (12) and (14) are identically satisfied and
the remaining equations turn out to be
\begin{eqnarray}
T_{0}\xi^0_{,n}=0,\quad (n=0,1,2),\\
T_{0}\xi^0_{,3}+T_{2}f\xi^3_{,0}=0,\\
T_{2}\xi^3_{,i}=0,\quad (i=1,2,3).
\end{eqnarray}
These equations yield the following four possibilities:
\begin{eqnarray*}
&(a)&\quad T_0=0,\quad T_{2}=0,\quad (b)\quad T_{0}=0,\quad T_{2}\neq0,\\
&(c)&\quad T_{0}\neq0,\quad T_{2}=0,\quad(d)\quad T_{0}\neq0,\quad
T_{2}\neq0.
\end{eqnarray*}
For the first possibility, we obtain
$\xi^0=\xi^0(x^a),~\xi^3=\xi^3(x^a)$. The second option implies that
$\xi^0=\xi^0(x^a)$ and $\xi^3$ to be an arbitrary constant. In the
case, when $T_{0}\neq0,~T_{2}=0$, we have
$\xi^0=constant,~\xi^3=\xi^3(x^a)$. The last possibility yields
both $\xi^0$ and $\xi^3$ to be arbitrary constant.\\
\par\noindent
{\bf Case II(iv):} $\xi^a=(0,~\xi^1,~\xi^2,~0)$\\
\par\noindent
In this case, Eq.(10) is identically satisfied and Eq.(7) implies
that $T_0=constant$. Rest of the equations yield the following
constraints
\begin{eqnarray*}
&(a)&\quad T_1=0,\quad T_{2}=0,\quad (b)\quad T_{1}=0,\quad T_{2}\neq0,\\
&(c)&\quad T_{1}\neq0,\quad T_{2}=0,\quad(d)\quad T_{1}\neq0,\quad
T_{2}\neq0.
\end{eqnarray*}
For the first case, MC equations yield
$\xi^1=\xi^1(x^a),~\xi^2=\xi^2(x^a)$. In the second possibility, we
obtain $\xi^1=\xi^1(x^a)$ and $\xi^2=cf$. The third option gives the
following solution $\xi^1=\frac{c}{\sqrt{T_1}}$ and
$\xi^2=\xi^2(x^a)$. In the last possibility, Eqs.(8), (9) and (13)
respectively imply that $\xi^1=\xi^1(r,\theta)$ and
$\xi^2=\xi^2(r,\theta)$. It follows from Eq.(11) that
$\xi^1=\frac{A(\theta)}{\sqrt{T_1}}$. Using this value of $\xi^1$ in
Eqs.(14) and (16) and combining them, we obtain
\begin{eqnarray}
\xi^2_{,2}-\frac{f_{,2}}{f}\xi^2=0
\end{eqnarray}
which gives $\xi^2=B(r)f$. Now we make use of $\xi^1,~\xi^2$ in
Eq.(9), it turns out that
\begin{eqnarray}
\frac{A_{,2}}{f}=-\frac{T_2}{\sqrt{T_1}}B'=\alpha,
\end{eqnarray}
where $\alpha$ is a constant which can be zero or non-zero. When
$\alpha=0$, we obtain
\begin{eqnarray}
\xi_{(1)}=\frac{1}{\sqrt{T_1}}\frac{\partial}{\partial
r},\quad\xi_{(2)}=f\frac{\partial}{\partial \theta}.
\end{eqnarray}
For $\alpha\neq0$, it follows that
\begin{eqnarray}
\xi^1=\frac{c_0+\alpha\int{fd\theta}}{\sqrt{T_1}},\quad\xi^2=(c_1-\alpha
\int{\frac{T_2}{\sqrt{T_1}}}dr)f
\end{eqnarray}
\par\noindent
{\bf Case II(v):} $\xi^a=(0,~\xi^1,~0,~\xi^3)$\\
\par\noindent
It follows from Eqs.(7) and (14) that $T_0=constant=T_2$. From
Eqs.(8), (12) and (10), (15) imply that $\xi^1$ and $\xi^3$ are
functions of $r$ and $\phi$ respectively. When we make use of
Eqs.(11), (13) and (15), it turns out that
$\xi^1=\frac{c}{\sqrt{T_1}}$ and $\xi^3$ becomes $constant$. Thus
the two MCs will be
\begin{eqnarray*}
\xi_{(1)}=\frac{1}{\sqrt{T_1}}\partial_r,\quad
\xi_{(2)}=\partial_\phi.
\end{eqnarray*}
\par\noindent
{\bf Case II(vi):} $\xi^a=(0,~0,~\xi^2,~\xi^3)$\\
\par\noindent
In this case, MC Eqs.(7)-(16) reduce to
\begin{eqnarray}
T_2\xi^2_{,m}=0,\quad T_2\xi^3_{,p}=0,\quad (m=0,1,2),\quad
(p=0,1),\\
T_{2}(\xi^2_{,3}+f^2\xi^3_{,2})=0,\\
T_2(\frac{f^2_{,2}}{f^2}\xi^2+2\xi^3_{,3})=0.
\end{eqnarray}
These imply that either $T_2=0$ or $T_2\neq0$. For the first option,
Eq.(37)-(39) give $\xi^2=\xi^2(x^a),~\xi^3=\xi^3(x^a)$. In the
second possibility, it follows from Eq.(37) that $\xi^2=\xi^2(\phi)$
and $\xi^3=\xi^3(\theta,\phi)$. Using Eqs.(38) and (39), it can be
shown, after some algebra, that
\begin{eqnarray}
\xi_{(1)}&=&\frac{2f^2}{f^2_{,2}}\sin\phi
exp(-2\int{\frac{1}{f^2_{,2}}d\theta})
\partial_\theta+\cos\phi exp(-2\int{\frac{1}{f^2_{,2}}d\theta})\partial_\phi,\nonumber\\
\xi_{(2)}&=&\frac{2f^2}{f^2_{,2}}\cos\phi
exp(-2\int{\frac{1}{f^2_{,2}}d\theta})\partial_\theta -\sin\phi
exp(-2\int{\frac{1}{f^2_{,2}}d\theta})\partial\phi.
\end{eqnarray}

\subsection{Three Components of $\xi^a(x^b)$ are Different from Zero}

It has four different possibilities:\\
III(i)\quad $\xi^a=(\xi^0,~\xi^1,~\xi^2,~0)$;\\
III(ii)\quad $\xi^a=(\xi^0,~\xi^1,~0,~\xi^3)$;\\
III(iii)\quad $\xi^a=(\xi^0,~0,~\xi^2,~\xi^3)$;\\
III(iv)\quad $\xi^a=(0,~\xi^1,~\xi^2,~\xi^3)$;\\
\par\noindent
{\bf Case III(i):} $\xi^a=(\xi^0,~\xi^1,~\xi^2,~0)$\\
\par\noindent
In this case, using Eqs.(10), (13) and (15) we find that
$\xi^0,~\xi^1$ and $\xi^2$ are functions of $t,r,\theta$ and from
Eq.(11), we have
\begin{eqnarray*}
\xi^1=\frac{A(t,\theta)}{\sqrt{|T_{1}|}},
\end{eqnarray*}
where $A(t,\theta)$ is an integration function. If we make use of
this value of $\xi^1$ in Eq.(16), we obtain
\begin{eqnarray*}
\xi^2=-\frac{T'_{2}}{2T_{2}\sqrt{|T_{1}|}}\frac{Af}{f_{,2}}.
\end{eqnarray*}
Substituting this value of $\xi^2$ together with $\xi^1$ in Eq.(14),
we obtain the value of $A$ as follows
\begin{equation}
A(t,\theta)=A_1(t)f_{,2},
\end{equation}
where $A_1(t)$ is an integration function. From Eq.(7), it follows
that
\begin{equation}
\xi^0=-\frac{T'_{0}}{2T_{0}\sqrt{|T_{1}|}}f_{,2}\int
A_1dt+B(r,\theta),
\end{equation}
where $B(r,\theta)$ is an integration function. Using values of
$\xi^1$ and $\xi^2$ in Eq.(12), we have
\begin{equation}
\frac{T_{2}}{\sqrt{|T_{1}|}}(\frac{T'_{2}}{2T_{2}\sqrt{|T_{1}|}})'
=\frac{f_{,22}}{f}=-k.
\end{equation}
Now plugging the values of $\xi^0$ and $\xi^1$ in Eq.(8), we obtain
\begin{equation}
\frac{T_{0}}{\sqrt{|T_{1}|}}(\frac{T'_{0}}{2T_{0}{\sqrt{|T_{1}|}}})'
=\frac{\ddot{A}_1}{A_1}=-\alpha^2,
\end{equation}
where $\alpha^2$ is a separation constant which may be positive,
negative or zero. Thus there arise six different possibilities:
\begin{eqnarray*}
&(a)&\quad \alpha^2>0,\quad k\neq 0;\quad(b)\quad \alpha^2<0,
\quad k\neq 0;\quad(c)\quad\alpha^2=0,\quad k\neq 0;\\
&(d)&\quad\alpha^2>0,\quad k=0;\quad(e)\quad\alpha^2<0,\quad k=0;
\quad(f)\quad\alpha^2=0,\quad k=0.
\end{eqnarray*}
For the case IIIi(a), after some algebraic manipulation, it is shown
that
\begin{eqnarray}
\xi^0&=&\frac{1}{\sqrt{|T_{1}|}}(c_0\cos\alpha t+c_1\sin\alpha
t)f_{,2},\nonumber\\
\xi^1&=&-\frac{f_{,2}}{\alpha}\frac{T'_{0}}{2
T_{0}\sqrt{|T_{1}|}}(c_0\sin\alpha t-c_1\cos\alpha
t)+c_2,\nonumber\\
\xi^2&=&-\frac{T'_{2}}{2 T_{2}\sqrt{|T_{1}|}}(c_0\cos\alpha
t+c_1\sin\alpha t)f,\quad T_{2}=\frac{k}{\alpha^2}T_{0}+c,
\end{eqnarray}
where $c$ is an arbitrary constants and $k$ can take values $\pm1$.
It follows that MCs are three which can be written as
\begin{eqnarray}
\xi_{(1)}&=&\frac{f_{,2}}{\sqrt{|T_{1}|}}\cos\alpha t\partial_t
-\frac{f_{,2}}{\alpha}\frac{T'_{0}}{2 T_{0}\sqrt{|T_{1}|}}\sin\alpha
t\partial_r-\frac{fT'_{2}}{2 T_{2}\sqrt{|T_{0}|}}\cos\alpha
t\partial_\theta,\nonumber\\
\xi_{(2)}&=&\frac{f_{,2}}{\sqrt{|T_{1}|}}\sin\alpha t\partial_t
+\frac{f_{,2}}{\alpha}\frac{T'_{0}}{2 T_{0}\sqrt{|T_{1}|}}\cos\alpha
t\partial_r-\frac{fT'_{2}}{2 T_{2}\sqrt{|T_{1}|}}\sin\alpha
t\partial_\theta,\nonumber\\
\xi_{(3)}&=&\partial_r.
\end{eqnarray}
In the case IIIi(b), $\alpha^2$ is replaced by $-\alpha^2$ in
Eq.(40) and it follows that
\begin{eqnarray}
\xi_{(1)}&=&\frac{f_{,2}}{\sqrt{|T_{1}|}}\cosh\alpha t\partial_t
-\frac{f_{,2}}{\alpha}\frac{T'_{0}}{2
T_{0}\sqrt{|T_{1}|}}\sinh\alpha t\partial_r-\frac{fT'_{2}}{2
T_{2}\sqrt{|T_{0}|}}\cosh\alpha
t\partial_\theta,\nonumber\\
\xi_{(2)}&=&\frac{f_{,2}}{\sqrt{|T_{1}|}}\sinh\alpha
t\partial_t-\frac{f_{,2}}{\alpha}\frac{T'_{0}}{2
T_{0}\sqrt{|T_{1}|}}\cosh\alpha t\partial_r-\frac{fT'_{2}}{2
T_{2}\sqrt{|T_{1}|}}\sinh\alpha
t\partial_\theta,\nonumber\\
\xi_{(3)}&=&\partial_r.
\end{eqnarray}
For the third subcase, we have $A_1=c_0t+c_1$ and
$\frac{T'_{0}}{2T_{0}\sqrt{|T_{1}|}}=\beta$, where $\beta$ is an
arbitrary constant. This implies that either $\beta$ is non-zero or
zero. If $\beta\neq 0$, it reduces to the case I(ii). For $\beta=0$,
we have $T_{0}=constant$. Using these values in Eq.(15) and (18) we
obtain
\begin{eqnarray}
\xi_{(1)}&=&f_{,2}\frac{1}{\sqrt{|T_{1}|}}t\partial_t
-\frac{f_{,2}T'_{2}}{2k T_{0} \sqrt{|T_{1}|}}\partial_r
-\frac{T'_{2}f}{2T_{2}\sqrt{|T_{1}|}}t\partial_\theta,\nonumber\\
\xi_{(2)}&=&\frac{1}{\sqrt{|T_{1}|}}f_{,2}\partial_t-\frac{T'_{2}f}{2T_{2}
\sqrt{|T_{1}|}}\partial_\theta,\nonumber\\
\xi_{(3)}&=&\partial_r,\nonumber\\
T_{0}=constant,\quad T'_{2}&=&2k\sqrt{|T_{1}|}\int\sqrt{|T_{1}|}dt.
\end{eqnarray}
The fourth and fifth subcases reduce to I(ii).

In the case Ii(f), we obtain $A_1=c_0t+c_1,\quad
\frac{T'_{0}}{2T_{0}\sqrt{|T_{1}|}}=\beta$ and $
\frac{T'_{2}}{2T_{2}\sqrt{|T_{1}|}}=\gamma$, where $\beta$ and
$\gamma$ are arbitrary constants. We see from Eq.(18) that $\gamma
T_{2}=0$ which implies that either $\gamma=0$ or $T_{2}=0$. This
gives rise to the following three possibilities
\begin{eqnarray*}
&(1)&\quad\gamma\neq 0\neq\beta,\quad T_{2}=0,\quad
(2)\quad\gamma=0,\quad\beta\neq 0\neq T_{2}\\
&(3)&\quad\gamma=0=\beta,\quad T_{2}\neq 0.
\end{eqnarray*}
In the subcase IIIif(1), for $k=0$, we have
\begin{eqnarray}
\xi_{(1)}&=&\frac{1}{\sqrt{|T_{1}|}}t\partial_t
+(\beta\frac{t^2}{2}+\int\frac{\sqrt{|T_{1}|}}{T_{0}}dr)
\partial_r
-\gamma\theta t\partial_\theta,\nonumber\\
\xi_{(2)}&=&\frac{1}{\sqrt{|T_{1}|}}\partial_t
-\beta t\partial_\theta,\nonumber\\
\xi_{(3)}&=&\partial_r.
\end{eqnarray}
For $k=\pm 1$, $c_0=0$. The subcases IIIif(2) and IIIif(3) reduce to
the case II(i). In the former subcase $T_{0}$ becomes $constant$
while in the latter subcase both $T_{0}$ and $T_{2}$ become
$constants$.\\
\par\noindent
{\bf Case III(ii):} $\xi^a=(\xi^0,~\xi^1,~0,~\xi^3)$\\
\par\noindent
In this case, when we replace $\xi^2=0$ in the MC Eqs.(7)-(16), it
follows eight different possibilities according to the values of
$T_0,~T_1,~T_2$. If we take at least one of these three components
of the energy-momentum tensor zero, we obtain infinite dimensional
MCs. The only case which gives finite dimensional MCs is that where
all $T_0,~T_1,~T_2$ are non-zero. Here we give the solution of this
case only. Solving MC equations (7)-(16) under the constraints of
this case, we obtain
\begin{equation}
\xi^0=\dot{A}\int{\frac{\sqrt{T_1}}{T_0}}dr+c_0, \quad
\xi^1=\frac{A}{\sqrt{T_1}},\quad \xi^3=c_1
\end{equation}
with
\begin{eqnarray*}
\frac{\ddot{A}}{A}=\frac{T'_{0}\sqrt{T_1}}{2T_0}
\int{\frac{\sqrt{T_1}}{T_0}}dr=\alpha,
\end{eqnarray*}
where $\alpha$ is a separation constant which gives three different
possibilities according to the value of $\alpha$ whether positive,
zero or negative. For $\alpha>0$, we have the following MCs
\begin{eqnarray}
\xi_{(1)}&=&\partial_t,\nonumber\\
\xi_{(2)}&=&\sqrt{\alpha} \exp(-\sqrt{\alpha}t)\partial_t
+\frac{\exp(-\sqrt{\alpha}t)}{\sqrt{T_1}}\partial_r,\nonumber\\
\xi_{(3)}&=&\exp(\sqrt{\alpha}t)
\int{\frac{\sqrt{T_1}}{T_0}}dr\partial_t
-\frac{\exp(\sqrt{\alpha}t)}{\sqrt{T_1}}\partial_r,\nonumber\\
\xi_{(4)}&=&\partial_\phi.
\end{eqnarray}
When $\alpha=0$, MCs turn out to be
\begin{eqnarray}
\xi_{(1)}&=&\partial_t,\nonumber\\
\xi_{(2)}&=&\int{\frac{\sqrt{T_1}}{T_0}}dr\partial_t
+\frac{t}{\sqrt{T_1}}\partial_r,\nonumber\\
\xi_{(3)}&=&\frac{1}{\sqrt{T_1}}\partial_r,\nonumber\\
\xi_{(4)}&=&\partial_\phi.
\end{eqnarray}
For $\alpha<0$, we obtain the following MCs
\begin{eqnarray}
\xi_{(1)}&=&\partial_t,\nonumber\\
\xi_{(2)}&=&\sqrt{\alpha} \sin(\sqrt{\alpha}t)\partial_t
+\frac{\cos(\sqrt{\alpha}t)}{\sqrt{T_1}}\partial_r,\nonumber\\
\xi_{(3)}&=&\cos(\sqrt{\alpha}t)
\int{\frac{\sqrt{T_1}}{T_0}}dr\partial_t
-\frac{\sin(\sqrt{\alpha}t)}{\sqrt{T_1}}\partial_r,\nonumber\\
\xi_{(4)}&=&\partial_\phi.
\end{eqnarray}
We see that in each case MCs turn out to be four.\\
\par\noindent
{\bf Case III(iii):} $\xi^a=(\xi^0,~0,~\xi^2,~\xi^3)$\\
\par\noindent
When we substitute $\xi^1=0$ in MC equations, we obtain the
following three cases.
\begin{eqnarray*}
(a)\quad T_0=0,\quad T_2\neq0,\quad(b)\quad T_0\neq0,\quad
T_2=0,\quad(c)\quad T_0\neq0,\quad T_2\neq0.
\end{eqnarray*}
The first two cases yields infinite dimensional MCs and the third
case is the interesting one which gives finite MCs. When we solve MC
equations simultaneously for this case, after some algebra, we
obtain the following MCs
\begin{eqnarray}
\xi_{(1)}&=&\partial_t,\nonumber\\
\xi_{(2)}&=&\frac{T_2}{T_0}f^2\cos\phi
\exp(-2\int{\frac{d\theta}{f^2_{,2}}})\partial_t
-t\cos\phi\partial_\theta+t\sin\phi
exp(-2\int{\frac{d\theta}{f^2_{,2}}})\partial_\phi,\nonumber\\
\xi_{(3)}&=&\frac{T_2}{T_0}f^2\sin\phi
\exp(-2\int{\frac{d\theta}{f^2_{,2}}})\partial_t
-t\sin\phi\partial_\theta-t\cos\phi
\partial_\phi,\nonumber\\
\xi_{(4)}&=&\cos\phi\partial_\theta-\sin\phi
exp(-2\int{\frac{d\theta}{f^2_{,2}}})\partial_\phi,\nonumber\\
\xi_{(5)}&=&\sin\phi\partial_\theta +\cos\phi\partial_\phi.
\end{eqnarray}
This gives five independent MCs.\\
\par\noindent
{\bf Case III(iv):} $\xi^a=(0,~\xi^1,~\xi^2,~\xi^3)$\\
\par\noindent
In this case, MC equations give the following four
possibilities:
\begin{eqnarray*}
(a)\quad T_1&=&0,\quad T_2\neq0,\quad(b)\quad T_1\neq0,\quad
T_2=0,\\
(c)\quad T_1&=&0,\quad T_2\neq0,\quad (d)\quad T_1\neq0,\quad
T_2\neq0.
\end{eqnarray*}
The first two cases reduce to the earlier one. For the case
IIIiv(c), we are left with the MC equations (14)-(16) along with
$\xi^1=\xi^1(r,\theta,\phi),~\xi^2=\xi^2(\theta,\phi),~\xi^3=\xi^3(\theta,\phi)$.
For $k=-1$, when we solve MC equations (14)-(16) we obtain the
following MCs
\begin{eqnarray}
\xi_{(1)}&=&\frac{2T_2}{T'_{2}}\cos\phi\sinh\theta\partial_r
-\cos\phi\cosh\theta\partial_\theta
+\sin\phi cosech\theta\partial_\phi,\nonumber\\
\xi_{(2)}&=&\frac{2T_2}{T'_{2}}\sin\phi\sinh\theta\partial_r
-\sin\phi\cosh\theta\partial_\theta
-\cos\phi cosech\theta\partial_\phi,\nonumber\\
\xi_{(3)}&=&\cos\phi\partial_\theta
-\sin\phi \coth\theta\partial_\phi,\nonumber\\
\xi_{(4)}&=&\sin\phi\partial_\theta
+\cos\phi \coth\theta\partial_\phi,\nonumber\\
\xi_{(5)}&=&\partial_\phi.
\end{eqnarray}
When $k=0$, MCs are
\begin{eqnarray}
\xi_{(1)}&=&\frac{2T_2}{T'_{2}}\cos\phi\theta\partial_r
-\cos\phi\frac{\theta^2}{2}\partial_\theta
-\sin\phi\frac{\theta}{2}\partial_\phi,\nonumber\\
\xi_{(2)}&=&\frac{2T_2}{T'_{2}}\sin\phi\theta\partial_r
-\sin\phi\frac{\theta^2}{2}\partial_\theta
+\cos\phi\frac{\theta}{2}\partial_\phi,\nonumber\\
\xi_{(3)}&=&\cos\phi\partial_\theta
-\sin\phi\frac{1}{\theta}\partial_\phi,\nonumber\\
\xi_{(4)}&=&\sin\phi\partial_\theta
-\cos\phi\frac{1}{\theta}\partial_\phi,\nonumber\\
\xi_{(5)}&=&\partial_\phi.
\end{eqnarray}
If we take $k=1$, MCs turn out to be
\begin{eqnarray}
\xi_{(1)}&=&\frac{2T_2}{T'_{2}}\cos\phi\sin\theta\partial_r
-\cos\phi\cos\theta\partial_\theta
+\sin\phi cosec\theta\partial_\phi,\nonumber\\
\xi_{(2)}&=&\frac{2T_2}{T'_{2}}\sin\phi\sin\theta\partial_r
-\sin\phi\cos\theta\partial_\theta
-\cos\phi cosec\theta\partial_\phi,\nonumber\\
\xi_{(3)}&=&\cos\phi\partial_\theta
-\sin\phi \cot\theta\partial_\phi,\nonumber\\
\xi_{(4)}&=&\sin\phi\partial_\theta
+\cos\phi \cot\theta\partial_\phi,\nonumber\\
\xi_{(5)}&=&\partial_\phi.
\end{eqnarray}
The case IIIiv(d) yields $\xi^1,~\xi^2,~\xi^3$ as functions of
$r,\theta,\phi$ with MC equations (14)-(16). For $k=-1$, when we
solve MC equations (14)-(16) we obtain the following MCs
\begin{eqnarray}
\xi_{(1)}&=&a[\frac{2T_2}{T'_{2}}(\cos\phi+\sin\phi)\sinh\theta
\partial_r-(\cos\phi+\sin\phi)\cosh\theta
\partial_\theta\nonumber\\
&-&(\sin\phi-\cos\phi) cosech\theta\partial_\phi],\nonumber\\
\xi_{(2)}&=&\cos\phi\partial_\theta
-\sin\phi \coth\theta\partial_\phi,\nonumber\\
\xi_{(3)}&=&\sin\phi\partial_\theta
+\cos\phi \coth\theta\partial_\phi,\nonumber\\
\xi_{(4)}&=&\partial_\phi.
\end{eqnarray}
When $k=0$, MCs take the following form
\begin{eqnarray}
\xi_{(1)}&=&a[\frac{2T_2}{T'_{2}}(\cos\phi+\sin\phi)\theta
\partial_r-(\cos\phi+\sin\phi)\frac{\theta^2}{2}
\partial_\theta\nonumber\\
&-&(\sin\phi-\cos\phi)\frac{\theta}{2}\partial_\phi],\nonumber\\
\xi_{(2)}&=&\cos\phi\partial_\theta
-\sin\phi\frac{1}{\theta}\partial_\phi,\nonumber\\
\xi_{(3)}&=&\sin\phi\partial_\theta
+\cos\phi\frac{1}{\theta}\partial_\phi,\nonumber\\
\xi_{(4)}&=&\partial_\phi.
\end{eqnarray}
If we take $k=1$, MCs turn out to be
\begin{eqnarray}
\xi_{(1)}&=&a[\frac{2T_2}{T'_{2}}(\cos\phi+\sin\phi)\sin\theta
\partial_r-(\cos\phi+\sin\phi)\cos\theta
\partial_\theta\nonumber\\
&-&(\sin\phi-\cos\phi) cosec\theta\partial_\phi],\nonumber\\
\xi_{(2)}&=&\cos\phi\partial_\theta
-\sin\phi \cot\theta\partial_\phi,\nonumber\\
\xi_{(3)}&=&\sin\phi\partial_\theta
+\cos\phi \cot\theta\partial_\phi,\nonumber\\
\xi_{(4)}&=&\partial_\phi.
\end{eqnarray}
This gives four independent MCs.

\subsection{Four Components of $\xi^a(x^b)$ are Different from Zero}

This is an interesting and a bit difficult case. In this section, we
shall evaluate MCs when all the four components of $\xi^a(x^b)$ are
non-zero. In other words, we find MCs only for those cases which
have non-degenerate energy-momentum tensor, i.e., $\det(T_{ab})\neq
0$. To this end, we set up the general conditions for the solution
of MC equations for the non-degenerate case.

When we solve Eqs.(7)-(16) simultaneously, after some tedious
algebra, we get the following solution
\begin{eqnarray}
\xi^0&=&\frac{T_{2}}{T_{0}}f^2[(\dot{A}_1(t,r)\sin\phi-\dot{A}_2(t,r)\cos\phi)
\int{\frac{1}{f^2}(\int{fd\theta})}d\theta]\nonumber\\
&-&\frac{T_{2}}{T_{1}}\dot{A}_3(t,r)\int{fd\theta}+A_4(t,r),
\end{eqnarray}
\begin{eqnarray}
\xi^1&=&\frac{T_{2}}{T_{0}}f^2[(A'_1(t,r)\sin\phi-A'_2(t,r)\cos\phi)
\int{\frac{1}{f^2}(\int{fd\theta})}d\theta]\nonumber\\
&-&\frac{T_{2}}{T_{1}}A'_3(t,r)\int{fd\theta}+A_5(t,r),
\end{eqnarray}
\begin{eqnarray}
\xi^2&=&[A_1(t,r)\sin\phi-A_2(t,r)\cos\phi]\int{fd\theta}
+c_1\sin\phi-c_2\cos\phi\nonumber\\
&+&A_3(t,r)f,
\end{eqnarray}
\begin{eqnarray}
\xi^3&=&-[(A_1(t,r)\cos\phi+A_2(t,r)\sin\phi)]
\int{\frac{1}{f^2}(\int{fd\theta})}d\theta\nonumber\\
&-&(c_1\sin\phi-c_2\cos\phi)\int{\frac{1}{f^2}d\theta}+c_3,
\end{eqnarray}
where $c_1,c_2,c_3$ are arbitrary constants and
$A_\nu=A_\nu(t,r),~\nu=1,2,3,4,5$. Here dot and prime indicate the
differentiation with respect to time  and $r$ coordinate
respectively. When we replace these values of $\xi^a$ in MC
Eqs.(7)-(16), we obtain the following differential constraints on
$A_\nu$ with $c_4=0$
\begin{eqnarray}
2 T_1 \ddot{A}_i+T_{0,1}A'_i=0,\quad i = 1,2,3,\\
2T_0 \dot{A_4}+T_{0,1}A_5 =0,\\
2T_2 \dot{A'_i} + T_{0} \left({T_{2} \over T_{0} } \right)'
\dot{A_i}= 0,\\
T_{0} A'_4 + T_{1} \dot{A_5} = 0,\\
\left\{T_{1,1} {T_2 \over T_1} + 2 T_1 \left({ T_2
\over T_1} \right)' \right\} A'_i + 2 T_2 A_i^{''} = 0,\\
T_{1,1} A_5 + 2 T_1 A'_5 = 0,\\
T_{2,1} A'_i + 2 T_1 A_i = 0,\quad c_0 = 0,\\
T_{2,1} A_5 = 0.
\end{eqnarray}
Thus the problem of working out MCs for all possibilities of
$A_i,~A_4,~A_5$ is reduced to solving the set of Eqs.(61)-(64)
subject to the above constraints. We start the classification of MCs
by considering the constraint Eq.(72). This can be satisfied for
three different possible cases
\begin{eqnarray*}
(i)\quad T'_{2} = 0,\quad A_5 \neq 0,\quad (ii)\quad T'_{2} \neq
0,\quad A_5 = 0,\quad (iii)\quad T'_{2}=0,\quad A_5 = 0.
\end{eqnarray*}
{\bf Case (i):} In this case, all the constraints remain unchanged
except (65), (69) and (71). Thus we have
\begin{eqnarray}
\dot{A'_i} - {1 \over 2} {T'_{0} \over T_{0} } \dot{A_i} = 0,\\
A''_i - {T'_{1} \over 2 T_{1} } A'_i = 0, \\
T_{1} A_i = 0.
\end{eqnarray}
The last equation is satisfied only if $ A_i = 0. $ As a result, all
the differential constraints involving $ A_i $ and its derivatives
disappear identically and we are left with Eqs.(66), (68) and (70)
only. Now integrating constraint Eq.(70) w.r.t. $ r $ and replacing
the value of $ A_5 $ in constraint Eq.(66), we have
$$ T'_{0} {A(t) \over \sqrt{T_{1}} } + 2T_{0} \dot{A_4} = 0,$$
where $ A(t) $ is an integration function. This gives rise to the
following two possibilities:
\par \noindent
\begin{eqnarray*}
(a)\quad T'_{0}=0,\quad\dot{A_4}=0,\quad(b)\quad T'_{0}\neq 0,
\quad\dot{A_4}\neq 0.
\end{eqnarray*}
For the case IVi(a), after some algebra, we arrive at the following
MCs
\begin{eqnarray}
\xi_{(1)}&=&\partial_t,\nonumber\\
\xi_{(2)}&=&{1\over \sqrt{T_{1}}}\partial_r,\nonumber\\
\xi_{(3)}&=&\cos\phi\partial_\theta
+\sin\phi\partial_\phi\int{\frac{d\theta}{f^2}},\nonumber\\
\xi_{(4)}&=&\sin\phi\partial_\theta
-\cos\phi\partial_\phi\int{\frac{d\theta}{f^2}},\nonumber\\
\xi_{(5)}&=&{1 \over a} \int \sqrt{T_{1}}dr\partial_t-{t\over
\sqrt{T_{1}} }\partial_r,\nonumber\\
\xi_{(6)}&=&\frac{\partial}{\partial \phi}.
\end{eqnarray}
This shows that we have six MCs.
\par \noindent
\par \noindent
In the case IVi(b), we have $\dot{A_4}\neq0$ and $T'_{0}\neq0$.
Solving Eqs.(66) and (68) and re-arranging terms, we get
\begin{equation}
{\ddot{A}\over A}={T_{0} \over 2\sqrt{T_{1}} }({ T'_{0} \over T_{0}
\sqrt{T_{1}} })'
 = \alpha,
\end{equation}
where $ \alpha $ is a separation constant and this gives the
following  three possible cases:
$$(1)\quad\alpha < 0,\quad(2)\quad\alpha = 0,\quad(3)\quad\alpha> 0. $$
\par \noindent
The first case $ \alpha < 0 $ reduces to the case IIIii(a) of the
previous section.
\par \noindent
The subcase IVib(2) gives
$$ A(t) = c_3 t + c_4 $$
and
\begin{equation}
{T'_{0} \over T_{0} \sqrt{T_{1}} } = \beta,
\end{equation}
where $\beta$ is an integration constant which yields the following
two possibilities
$$ (*)\quad \beta \neq 0,\quad(**)\quad \beta = 0. $$
\par \noindent
The first possibility implies that $$ T_{0} = \beta_0 e^{\beta \int
\sqrt{T_{1}} dr},$$ where $ \beta_0 $ is an integration constant.
Now we solve Eqs.(70) and (72) by using this constraint, we can get
the following MCs
\begin{eqnarray}
\xi_{(1)}&=&\partial_t,\nonumber\\
\xi_{(2)}&=&\sin \phi\partial_\theta
-\cos\phi\int{\frac{d\theta}{f^2}}\partial_\phi,\nonumber\\
\xi_{(3)}&=&\cos\phi\partial_\theta
+\sin\phi\int{\frac{d\theta}{f^2}}\partial_\phi,\nonumber\\
\xi_{(4)}&=&{1\over \beta_0} \int {\sqrt{T_{1}} \over e^{\beta \int
\sqrt{T_{1}} d r} }d r\partial_t-{t \over \sqrt{T_{1}}}\partial_r,\nonumber\\
\xi_{(5)}&=&{1\over \sqrt{T_{1}}}\partial_r.
\end{eqnarray}
This gives five independent MCs.
\par \noindent
\par \noindent
For the case IVib2(**), $T_0=constant$. Using this fact Eq.(70)
yields $A_4=g(r)$. Thus we have the solution
\begin{eqnarray}
\xi_{(1)}&=&\partial_t,\nonumber\\
\xi_{(2)}&=&\sin \phi\partial_\theta
-\cos\phi\int{\frac{d\theta}{f^2}}\partial_\phi,\nonumber\\
\xi_{(3)}&=&\cos\phi\partial_\theta
+\sin\phi\int{\frac{d\theta}{f^2}}\partial_\phi,\nonumber\\
\xi_{(4)}&=&\frac{1}{b}\int{\sqrt{T_1}}d r\partial_t
-{t \over \sqrt{T_{1}}}\partial_r,\nonumber\\
\xi_{(5)}&=&{1\over \sqrt{T_{1}}}\partial_r.
\end{eqnarray}
We again have five MCs.
\par \noindent
The case IVib(3) gives the same results as the case IVib(1).\\
\par \noindent
{\bf Case (ii):} In the case IV(ii), when $T'_2\neq 0$, it follows
from Eqs.(69)-(72) that for
$\frac{T_2}{\sqrt{T_1}}(\frac{T'_2}{2T_2\sqrt{T_1}})'+1\neq 0$, we
obtain the following four MCs
\begin{eqnarray}
\xi_{(1)}&=&\partial_t,\nonumber\\
\xi_{(2)}&=&\sin \phi\partial_\theta
-\cos\phi\int{\frac{d\theta}{f^2}}\partial_\phi,\nonumber\\
\xi_{(3)}&=&\cos\phi\partial_\theta
+\sin\phi\int{\frac{d\theta}{f^2}}\partial_\phi,\nonumber\\
\xi_{(4)}&=&\partial_\phi.
\end{eqnarray}
If $\frac{T_2}{\sqrt{T_1}}(\frac{T'_2}{2T_2\sqrt{T_1}})'+1=0$ and
$(\frac{T'_2}{\sqrt{T_0T_1T_2}})'\neq 0$, we have seven MCs given by
\begin{eqnarray}
\xi_{(1)}&=&\partial_t,\nonumber\\
\xi_{(2)}&=&\sin \phi\partial_\theta
-\cos\phi\int{\frac{d\theta}{f^2}}\partial_\phi,\nonumber\\
\xi_{(3)}&=&\cos\phi\partial_\theta
+\sin\phi\int{\frac{d\theta}{f^2}}\partial_\phi,\nonumber\\
\xi_{(4)}&=&\partial_\phi,\nonumber\\
\xi_{(5)}&=&\frac{1}{\sqrt{T_1}}\sin\phi\partial_r
f^2\int{(\frac{1}{f^2}}\int{fd\theta})d\theta
+X\sin\phi\partial_\theta\int{f}d\theta\nonumber\\
&+&X\cos\phi\partial_\phi\int{(\frac{1}{f^2}}\int{fd\theta})d\theta,\nonumber\\
\xi_{(6)}&=&\frac{1}{\sqrt{T_1}}\cos\phi\partial_r
f^2\int{(\frac{1}{f^2}}\int{fd\theta})d\theta
+X\cos\phi\partial_\theta\int{f}d\theta\nonumber\\
&+&X\sin\phi\partial_\phi\int{(\frac{1}{f^2}}\int{fd\theta})d\theta,\nonumber\\
\xi_{(7)}&=&\frac{1}{\sqrt{T_1}}\partial_r\int{f}d\theta
+Xf\sin\theta\partial_\theta.
\end{eqnarray}
where $X=\frac{T'_2}{2T_2\sqrt{T_1}}$. If we have
$\frac{T_2}{\sqrt{T_1}}(\frac{T'_2}{2T_2\sqrt{T_1}})'+1=0,~
(\frac{T'_2}{\sqrt{T_0T_1T_2}})'=0$ and $(\frac{T'_0}{T'_2})'\neq
0$, then we get the same MCs as given by Eq.(81).

When $\frac{T_2}{\sqrt{T_1}}(\frac{T'_2}{2T_2\sqrt{T_1}})'+1=0,~
(\frac{T'_2}{\sqrt{T_0T_1T_2}})'=0$ and $\frac{T'_0}{T'_2}=\delta$,
an arbitrary constant. For $\delta>0$, we obtain
\begin{eqnarray}
\xi_{(1)}&=&\partial_t,\nonumber\\
\xi_{(2)}&=&\sin \phi\partial_\theta
-\cos\phi\int{\frac{d\theta}{f^2}}\partial_\phi,\nonumber\\
\xi_{(3)}&=&\cos\phi\partial_\theta
+\sin\phi\int{\frac{d\theta}{f^2}}\partial_\phi,\nonumber\\
\xi_{(4)}&=&\partial_\phi,\nonumber\\
\xi_{(5)}&=&(\frac{T_2}{T_0}X\sqrt{\delta}\sinh\sqrt{\delta}t\partial_t
-\frac{1}{\sqrt{T_1}}\cosh\sqrt{\delta}t\partial_r)\sin\phi
f^2\int{(\frac{1}{f^2}}\int{fd\theta})d\theta\nonumber\\
&-&(\sin\phi\partial_\theta\int{fd\theta}+\cos\phi\partial_\phi
\int{(\frac{1}{f^2}}\int{fd\theta})d\theta)
X\cosh\sqrt{\delta}t,\nonumber\\
\xi_{(6)}&=&(\frac{T_2}{T_0}X\sqrt{\delta}\sinh\sqrt{\delta}t\partial_t
-\frac{1}{\sqrt{T_1}}\cosh\sqrt{\delta}t\partial_r)\cos\phi
f^2\int{(\frac{1}{f^2}}\int{fd\theta})d\theta\nonumber\\
&-&(\cos\phi\partial_\theta\int{fd\theta}-\sin\phi\partial_\phi
\int{(\frac{1}{f^2}}\int{fd\theta})d\theta))
X\cosh\sqrt{\delta}t,\nonumber\\
\xi_{(7)}&=&(\frac{T_2}{T_0}X\sqrt{\delta}\sinh\sqrt{\delta}t\partial_t
-\frac{1}{\sqrt{T_1}}\cosh\sqrt{\delta}t\partial_r)\int{fd\theta}\nonumber\\
&+&X\cosh\sqrt{\delta}t\partial_\theta
f^2\int{(\frac{1}{f^2}}\int{fd\theta})d\theta,\nonumber\\
\xi_{(8)}&=&(\frac{T_2}{T_0}X\sqrt{\delta}\cosh\sqrt{\delta}t\partial_t
-\frac{1}{\sqrt{T_1}}\sinh\sqrt{\delta}t\partial_r)\sin\phi
f^2\int{(\frac{1}{f^2}}\int{fd\theta})d\theta\nonumber\\
&-&(\sin\phi\partial_\theta\int{fd\theta}+\cos\phi\partial_\phi
\int{(\frac{1}{f^2}}\int{fd\theta})d\theta)
X\sinh\sqrt{\delta}t,\nonumber\\
\xi_{(9)}&=&(\frac{T_2}{T_0}X\sqrt{\delta}\cosh\sqrt{\delta}t\partial_t
-\frac{1}{\sqrt{T_1}}\sinh\sqrt{\delta}t\partial_r)\cos\phi
f^2\int{(\frac{1}{f^2}}\int{fd\theta})d\theta\nonumber\\
&-&(\cos\phi\partial_\theta\int{fd\theta}-\sin\phi\partial_\phi
\int{(\frac{1}{f^2}}\int{fd\theta})d\theta)
X\sinh\sqrt{\delta}t,\nonumber\\
\xi_{(10)}&=&(\frac{T_2}{T_0}X\sqrt{\delta}\cosh\sqrt{\delta}t\partial_t
-\frac{1}{\sqrt{T_1}}X\sinh\sqrt{\delta}t\partial_r)\int{fd\theta}\nonumber\\
&+&X\sinh\sqrt{\delta}t\partial_\theta
f^2\int{(\frac{1}{f^2}}\int{fd\theta})d\theta.
\end{eqnarray}
If $\delta=0$, we have
\begin{eqnarray}
\xi_{(1)}&=&\partial_t,\nonumber\\
\xi_{(2)}&=&\sin \phi\partial_\theta
-\cos\phi\int{\frac{d\theta}{f^2}}\partial_\phi,\nonumber\\
\xi_{(3)}&=&\cos\phi\partial_\theta
+\sin\phi\int{\frac{d\theta}{f^2}}\partial_\phi,\nonumber\\
\xi_{(4)}&=&\partial_\phi,\nonumber\\
\xi_{(5)}&=&(\frac{T_2}{T_0}X\partial_t
-\frac{1}{\sqrt{T_1}}t\partial_r)\sin\phi
f^2\int{(\frac{1}{f^2}}\int{fd\theta})d\theta\nonumber\\
&-&(\sin\phi\partial_\theta\int{fd\theta}+\cos\phi\partial_\phi
\int{(\frac{1}{f^2}}\int{fd\theta})d\theta)Xt,\nonumber\\
\xi_{(6)}&=&(-\frac{T_2}{T_0}X\partial_t
+\frac{1}{\sqrt{T_1}}t\partial_r)\cos\phi
f^2\int{(\frac{1}{f^2}}\int{fd\theta})d\theta\nonumber\\
&+&(\cos\phi\partial_\theta\int{fd\theta}-\sin\phi\partial_\phi
\int{(\frac{1}{f^2}}\int{fd\theta})d\theta)Xt,\nonumber\\
\xi_{(7)}&=&(\frac{T_2}{T_0}X\partial_t
-\frac{1}{\sqrt{T_1}}t\partial_r)\int{fd\theta}
+Xt\partial_\theta f^2\int{(\frac{1}{f^2}}\int{fd\theta})d\theta,\nonumber\\
\xi_{(8)}&=&(\frac{1}{\sqrt{T_1}}\partial_r
f^2\int{(\frac{1}{f^2}}\int{fd\theta})d\theta
+X\partial_\theta\int{fd\theta})\sin\phi\nonumber\\
&+&X\cos\phi\partial_\phi
\int{(\frac{1}{f^2}}\int{fd\theta})d\theta,\nonumber\\
\xi_{(9)}&=&(\frac{1}{\sqrt{T_1}}\partial_rf^2
\int{(\frac{1}{f^2}}\int{fd\theta})d\theta
+X\partial_\theta\int{fd\theta})\cos\phi\nonumber\\
&-&X\sin\phi\partial_\phi
\int{(\frac{1}{f^2}}\int{fd\theta})d\theta,\nonumber\\
\xi_{(10)}&=&\frac{1}{\sqrt{T_1}}\partial_r\int{fd\theta}-X\partial_\theta
f^2\int{(\frac{1}{f^2}}\int{fd\theta})d\theta.
\end{eqnarray}
For $\delta<0$, MCs are given by
\begin{eqnarray}
\xi_{(1)}&=&\partial_t,\nonumber\\
\xi_{(2)}&=&\sin \phi\partial_\theta
-\cos\phi\int{\frac{d\theta}{f^2}}\partial_\phi,\nonumber\\
\xi_{(3)}&=&\cos\phi\partial_\theta
+\sin\phi\int{\frac{d\theta}{f^2}}\partial_\phi,\nonumber\\
\xi_{(4)}&=&\partial_\phi,\nonumber\\
\xi_{(5)}&=&(-\frac{T_2}{T_0}X\sqrt{-\delta}\sin\sqrt{-\delta}t\partial_t
-\frac{1}{\sqrt{T_1}}\cos\sqrt{-\delta}t\partial_r)\sin\phi f^2
\int{(\frac{1}{f^2}}\int{fd\theta})d\theta\nonumber\\
&-&(\sin\phi\partial_\theta\int{fd\theta}+\cos\phi\partial_\phi
\int{(\frac{1}{f^2}}\int{fd\theta})d\theta))
X\cos\sqrt{-\delta}t,\nonumber\\
\xi_{(6)}&=&(\frac{T_2}{T_0}X\sqrt{-\delta}\sin\sqrt{-\delta}t\partial_t
+\frac{1}{\sqrt{T_1}}\cos\sqrt{-\delta}t\partial_r)\cos\phi
f^2\int{(\frac{1}{f^2}}\int{fd\theta})d\theta\nonumber\\
&+&(\cos\phi\partial_\theta\int{fd\theta}-\sin\phi\partial_\phi
f^2\int{(\frac{1}{f^2}}\int{fd\theta})d\theta))
X\cos\sqrt{-\delta}t,\nonumber\\
\xi_{(7)}&=&(\frac{T_2}{T_0}X\sqrt{-\delta}\sin\sqrt{-\delta}t\partial_t
+\frac{1}{\sqrt{T_1}}\cos\sqrt{-\delta}t\partial_r)\int{fd\theta}\nonumber\\
&-&X\cos\sqrt{-\delta}t\partial_\theta f^2
\int{(\frac{1}{f^2}}\int{fd\theta})d\theta,\nonumber\\
\xi_{(8)}&=&(\frac{T_2}{T_0}X\sqrt{-\delta}\cos\sqrt{-\delta}t\partial_t
-\frac{1}{\sqrt{T_1}}\sin\sqrt{-\delta}t\partial_r)\sin\phi
f^2\int{(\frac{1}{f^2}}\int{fd\theta})d\theta)\nonumber\\
&-&(\sin\phi\partial_\theta\int{fd\theta}+\cos\phi\partial_\phi
\int{(\frac{1}{f^2}}\int{fd\theta})d\theta))
X\sin\sqrt{-\delta}t,\nonumber\\
\xi_{(9)}&=&(\frac{T_2}{T_0}X\sqrt{\delta}\cos\sqrt{-\delta}t\partial_t
-\frac{1}{\sqrt{T_1}}\sin\sqrt{-\delta}t\partial_r)\cos\phi
f^2\int{(\frac{1}{f^2}}\int{fd\theta})d\theta\nonumber\\
&-&(\cos\phi\partial_\theta\int{fd\theta}-\sin\phi\partial_\phi
f^2\int{(\frac{1}{f^2}}\int{fd\theta})d\theta)
X\sin\sqrt{-\delta}t,\nonumber\\
\xi_{(10)}&=&(\frac{T_2}{T_0}X\sqrt{-\delta}\cos\sqrt{-\delta}t\partial_t
-\frac{1}{\sqrt{T_1}}X\sin\sqrt{-\delta}t\partial_r)\int{fd\theta}\nonumber\\
&+&X\sin\sqrt{-\delta}t\partial_\theta
f^2\int{(\frac{1}{f^2}}\int{fd\theta})d\theta.
\end{eqnarray}
From Eqs.(83)-(85), it follows that for each value of $\delta$, we
obtain ten independent MCs.
\par \noindent
{\bf Case (iii):} In this case, we have $T_2=constant$ and $A_5= 0$.
This can be solved trivially and gives similar results as in the
case IVib(1).

\section{Conclusion}

A large part of GR research is the consequence of classifying
solutions of the EFEs. There are various approaches to classify
spacetimes, including the Segre classification of the
energy-momentum tensor or the Petrov classification of the Weyl
tensor and have been studied extensively by many researchers [23].
They also classify spacetimes using symmetry vector fields, in
particular KVs and HVs. KVs may be used to classify spacetimes as
there is a limit to the number of global, smooth Killing vector
fields that a spacetime may possess (the maximum being 10 for
4-dimensional spacetimes). Generally speaking, the higher the
dimension of the algebra of symmetry vector fields on a spacetime,
the more symmetry the spacetime admits. For example, the
Schwarzschild solution has a Killing algebra of dimension 4 (3
spatial rotational vector fields and a time translation), whereas
the Friedmann-Robertson-Walker metric (excluding the Einstein static
subcase) has a Killing algebra of dimension 6 (3 translations and 3
rotations). The Einstein static metric has a Killing algebra of
dimension 7 (the previous 6 plus a time translation). The assumption
of a spacetime admitting a certain symmetry vector field can place
severe restrictions on the spacetime.

The relation between geometry and physics can be highlighted if the
vector field $\xi$ is regarded as preserving certain physical
quantities along the flow lines of $\xi$. Since every KV is a MC
hence for a given solution of the EFE, a vector field that preserves
the metric necessarily preserves the corresponding energy-momentum
tensor. When the energy-momentum tensor represents a perfect fluid,
every KV preserves the energy density, pressure and the fluid flow
vector field. When the energy-momentum tensor represents an
electromagnetic field, a KV does not necessarily preserve the
electric and magnetic fields. The main application of these
symmetries occur in GR, where solutions of EFEs may be classified by
imposing some certain symmetries on the spacetime.

This paper continues the study of the symmetries of the
energy-momentum tensor for the static spacetimes with maximal
symmetric transverse spaces. We have classified static spacetimes
with maximal symmetric transverse spaces according to their MCs. The
MC equations are solved by taking one non-zero component, two
non-zero components, three non-zero components and finally all
non-zero components. The case with all the non-zero components gives
the generalization of the static spherically symmetric spacetimes
and we recover all the results from the spacetime under
consideration. The results can be summarized in the form of tables
given below:

\vspace{0.1cm}

{\bf {\small Table 1}. }{\small MCs when one component of $\xi^a$ is
non-zero}.

\vspace{0.1cm}

\begin{center}
\begin{tabular}{|l|l|l|}
\hline {\bf Cases} & {\bf MCs} & {\bf Constraints}
\\ \hline Ii(a) & Infinite dimensional MCs & $T_0=0,~T_1\neq 0,~T_2\neq 0$
\\ \hline Ii(b) &$1$&
$T_0\neq0,~T_1\neq 0,~T_2\neq 0$
\\ \hline Iii(a) & Infinite dimensional MCs&
$T_0=constant=T_2,~T_1=0,~T'_1=0$
\\ \hline Iii(b) & $1$&$
\begin{array}{c}
T_0=constant=T_2,\\T_1=constant\neq 0
\end{array}
$
\\ \hline Iii(c) & $1$&$
\begin{array}{c}
T_0=constant=T_2,\\T'_1\neq0,~T_1\neq 0
\end{array}
$
\\ \hline I(iii) & Infinite dimensional MCs&
$T_0\neq0,~T_1\neq 0,~T_2=0$
\\ \hline Iiv(a) & Infinite dimensional MCs&
$T_0\neq0,~T_1\neq 0,~T_2=0$
\\ \hline Iiv(b) & $1$&
$T_0\neq0,~T_1\neq 0,~T_2\neq 0$
\\ \hline
\end{tabular}
\end{center}

\vspace{0.2cm}

{\bf {\small Table 2}. }{\small MCs when two components of $\xi^a$
are non-zero}.

\vspace{0.1cm}

\begin{center}
\begin{tabular}{|l|l|l|}
\hline {\bf Cases} & {\bf MCs} & {\bf Constraints}
\\ \hline IIi(a) & Infinite dimensional MCs&
$T_0=0=T_1,~T_2=constant$
\\ \hline IIi(b) & Infinite dimensional MCs&
$T_0=0,~T_1\neq0,~T_2=constant$
\\ \hline IIi(c) & Infinite dimensional MCs&
$T_0\neq0,~T_1=0,~T_2=constant$
\\ \hline IIid(1) & $3$&$
\begin{array}{c}
T_0\neq0,~T_1\neq0,~T_2=constant,\\
\frac{T_{0}}{\sqrt{|T_{1}|}}
(\frac{T'_{0}}{2T_{0}\sqrt{|T_{1}|}})'=-\alpha^2,~\alpha^2>0
\end{array}
$
\\ \hline IIid(2) & $3$&$
\begin{array}{c}
T_0\neq0,~T_1\neq0,~T_2=constant,\\
\alpha^2<0
\end{array}
$
\\ \hline IIid3(*) & $3$&$
\begin{array}{c}
T_0\neq0,~T_1\neq0,~T_2=constant,\\
\alpha^2=0,\\
\frac{T'_{0}}{2T_{0}\sqrt{|T_{1}|}}=\beta\neq0
\end{array}
$
\\ \hline IIid3(**) & $3$&$
\begin{array}{c}
T_0\neq0,~T_1\neq0,~T_2\neq 0,~\alpha^2=0,\\
\beta=0
\end{array}
$
\\ \hline IIii(a) & Infinite dimensional MCs&
$T_0=0,~T_1\neq0,~T_2\neq 0$
\\ \hline IIii(b) & $1$&
$T_0\neq0,~T_1\neq0,~T_2\neq 0$
\\ \hline IIiii(a) & Infinite dimensional MCs&
$T_0=0,~T_1\neq0,~T_2=0$
\\ \hline IIiii(b) & Infinite dimensional MCs&$
\begin{array}{c}
T_0=0,~T_1\neq0,~T_2\neq 0
\end{array}
$
\\ \hline
\end{tabular}
\end{center}
\begin{center}
\begin{tabular}{|l|l|l|}
\hline {\bf Cases} & {\bf RCs} & {\bf Constraints}
\\ \hline IIiii(c) & Infinite dimensional MCs&$
\begin{array}{c}
T_0\neq0,~T_1\neq0,~T_2=0
\end{array}
$
\\ \hline IIiii(d) & $2$&$
\begin{array}{c}
T_0\neq0,~T_1\neq0,~T_2\neq 0
\end{array}
$
\\ \hline
IIiv(a) & Infinite dimensional MCs& $T_0=constant,~T_1=0=T_2$
\\ \hline IIiv(b) & Infinite dimensional MCs&$
\begin{array}{c}
T_0=constant,~T_1=0,~T_2\neq 0
\end{array}
$
\\ \hline IIiv(c) & Infinite dimensional MCs&$
\begin{array}{c}
T_0=constant,~T_1\neq0,~T_2=0
\end{array}
$
\\ \hline IIivd(1) & $2$&$
\begin{array}{c}
T_0\neq0,~T_1\neq0,~T_2\neq 0,\\
-\frac{T_{2}}{\sqrt{T_{1}}}B'=\alpha,~\alpha=0
\end{array}
$
\\ \hline IIivd(2) & $2$&$
\begin{array}{c}
T_0\neq0,~T_1\neq0,~T_2\neq 0,~\alpha\neq0
\end{array}
$
\\ \hline II(v) & $2$&$
\begin{array}{c}
T_0=constant,~T_1\neq0,\\
T_2=constant
\end{array}
$
\\ \hline IIvi(a) & Infinite dimensional MCs &$
\begin{array}{c}
T_0\neq0,~T_1\neq0,~T_2=0
\end{array}
$
\\ \hline IIvi(b) & $2$&$
\begin{array}{c}
T_0\neq0,~T_1\neq0,~T_2\neq 0
\end{array}
$
\\ \hline
\end{tabular}
\end{center}
\vspace{0.2cm}

{\bf {\small Table 3}. }{\small MCs when three components of $\xi^a$
are non-zero}.

\vspace{0.1cm}

\begin{center}
\begin{tabular}{|l|l|l|}
\hline {\bf Cases} & {\bf MCs} & {\bf Constraints}
\\ \hline IIIi(a) & $3$&
$\begin{array}{c}
T_0\neq0,~T_1\neq0,~k\neq0\\
\frac{T_{0}}{\sqrt{|T_{1}|}}(\frac{T'_{0}}{2T_{0}{\sqrt{|T_{1}|}}})'
=-\alpha^2,~\alpha^2>0,\\
T_2=\frac{k}{\alpha^2}T_0+c
\end{array}
$
\\ \hline IIIi(b) & $3$&
$\begin{array}{c}
T_0\neq0,~T_1\neq0,~k\neq0,~\alpha^2<0,\\
T_2=\frac{k}{\alpha^2}T_0+c
\end{array}
$
\\ \hline IIIic(1) & I(ii)&
$\begin{array}{c}
T_0=constant,~T_1\neq0,~k\neq0,\\
\alpha^2=0,~T'_2=2k\sqrt{T_1}\int{\sqrt{T_1}}dt,\\
\frac{T'_0}{2T_0\sqrt{T_1}}=\beta,~\beta\neq0
\end{array}$
\\ \hline IIIic(2) & $3$&$
\begin{array}{c}
T_0=constant,~T_1\neq0,~k\neq0,\\
\alpha^2=0,~T'_2=2k\sqrt{T_1}\int{\sqrt{T_1}}dt,\\
\beta=0
\end{array}
$
\\ \hline IIIi(d) & II(i)&$
\begin{array}{c}
T_0\neq0,~T_1\neq0,~T_2\neq0,~k=0,\\
\alpha^2>0
\end{array}
$
\\ \hline IIIi(e) & II(i)&$
\begin{array}{c}
T_0\neq0,~T_1\neq0,~T_2\neq0,~k=0,\\
\alpha^2<0
\end{array}
$
\\ \hline
\end{tabular}
\end{center}
\begin{center}
\begin{tabular}{|l|l|l|}
\hline {\bf Cases} & {\bf RCs} & {\bf Constraints}
\\ \hline
IIIif(1) & $3$&$
\begin{array}{c}
T_0\neq0,~T_1\neq0,~T_2=0,~k=0,\\
\alpha^2=0,~\frac{T'_0}{2T_0\sqrt{T_1}}=\beta,~
\frac{T'_2}{2T_2\sqrt{T_1}}=\gamma,\\
\beta\neq0,~\gamma\neq0
\end{array}
$
\\ \hline IIIif(2) & I(ii)&$
\begin{array}{c}
T_0=constant,~T_1\neq0,~T_2\neq0,\\
k=0,~\beta\neq0,~\gamma=0
\end{array}
$
\\ \hline IIIif(3) & I(ii)&$
\begin{array}{c}
T_0=constant,~T_1\neq0,\\
T_2=constant,~k=0,~\beta=0=\gamma
\end{array}
$
\\ \hline IIIii(a) & Infinite dimensional MCs&$
\begin{array}{c}
T_0=0,~T_1=0,~T_2=0
\end{array}
$
\\ \hline IIIii(b) & Infinite dimensional MCs&$
\begin{array}{c}
T_0=0,~T_1=0,~T_2=constant\neq0
\end{array}
$
\\ \hline IIIii(c) & Infinite dimensional MCs&$
\begin{array}{c}
T_0=0,~T_1\neq0,~T_2=0
\end{array}
$
\\ \hline IIIii(d) & Infinite dimensional MCs&$
\begin{array}{c}
T_0\neq0,~T_1=0,~T_2=0
\end{array}
$
\\ \hline IIIii(e) & Infinite dimensional MCs&$
\begin{array}{c}
T_0=0,~T_1\neq0,~T_2=constant\neq0
\end{array}
$
\\ \hline IIIii(f) & Infinite dimensional MCs&$
\begin{array}{c}
T_0\neq0,~T_1\neq0,~T_2=0
\end{array}
$
\\ \hline IIIii(g) & Infinite dimensional MCs&$
\begin{array}{c}
T_0\neq0,~T_1=0,~T_2=constant\neq0
\end{array}
$
\\ \hline IIIiih(1) & $4$&$
\begin{array}{c}
T_0\neq0,~T_1\neq0,~T_2=constant\neq0,\\
\frac{T'_0\sqrt{T_1}}{2T_0}\int{\frac{\sqrt{T_1}}{T_0}}dr=\alpha,~
\alpha>0
\end{array}
$
\\ \hline IIIiih(2) & $4$&$
\begin{array}{c}
T_0\neq0,~T_1\neq0,~T_2=constant\neq0,\\
\alpha=0
\end{array}
$
\\ \hline IIIiih(3) & $4$&$
\begin{array}{c}
T_0\neq0,~T_1\neq0,~T_2=constant\neq0,\\
\alpha<0
\end{array}
$
\\ \hline IIIiii(a) & Infinite dimensional MCs&$
\begin{array}{c}
T_0=0,~T_1\neq0,~T_2\neq0
\end{array}
$
\\ \hline IIIiii(b) & Infinite dimensional MCs&$
\begin{array}{c}
T_0\neq0,~T_1\neq0,~T_2=0
\end{array}
$
\\ \hline IIIiii(c) & $5$&$
\begin{array}{c}
T_0\neq0,~T_1\neq0,~T_2\neq0
\end{array}
$
\\ \hline IIIiv(a) & Infinite dimensional MCs&$
\begin{array}{c}
T_0\neq0,~T_1=0,~T_2\neq0
\end{array}
$
\\ \hline IIIiv(b) & Infinite dimensional MCs&$
\begin{array}{c}
T_0\neq0,~T_1\neq0,~T_2=0
\end{array}
$
\\ \hline IIIiv(c) & $5$&$
\begin{array}{c}
T_0\neq0,~T_1=0,~T_2\neq0
\end{array}
$
\\ \hline IIIiv(d) & $4$&$
\begin{array}{c}
T_0\neq0,~T_1\neq0,~T_2\neq0
\end{array}
$
\\ \hline
\end{tabular}
\end{center}
\newpage

{\bf {\small Table 4}. }{\small MCs when all components of $\xi^a$
are non-zero}.

\vspace{0.1cm}

\begin{center}
\begin{tabular}{|l|l|l|}
\hline {\bf Cases} & {\bf MCs} & {\bf Constraints}
\\ \hline IVi(a) & $6$&
$\begin{array}{c} T'_2=0,~ A_5(t,r)\neq0,~T'_0=0,\\
\dot{A}_4(t,r)=0,
\end{array} $
\\ \hline IVib(1) & IIIii(a)&
$\begin{array}{c} T'_2\neq0,~
A_5(t,r)\neq0,~T'_0=0,\\
\dot{A}_4(t,r)\neq0,~\frac{T_0}{\sqrt{T_1}}(\frac{T'_0}{2T_0\sqrt{T_1}})'=\alpha,\\
\alpha<0
\end{array}
$
\\ \hline IVib2(*) & $5$&
$\begin{array}{c} T'_2\neq0,~
A_5(t,r)\neq0,~T'_0=0,\\
\alpha=0,~\frac{T'_0}{T_0\sqrt{T_1}}=\beta,~ \beta\neq0
\end{array}$
\\ \hline IVib2(**) & $5$&
$\begin{array}{c} T'_2\neq0,~
A_5(t,r)\neq0,~T'_0=0,\\
\alpha=0,~\beta=0
\end{array}
$
\\ \hline IVib(3) & IVib(1)&
$\begin{array}{c}
T'_2\neq0,~A_5(t,r)\neq0,~T'_0=0,\\
\alpha>0
\end{array}
$
\\ \hline IVii(a) & $4$&$
\begin{array}{c}
T'_2\neq0,~A_5(t,r)=0,~\frac{T_2}{\sqrt{T_1}}(\frac{T'_2}{2T_2\sqrt{T_1}})'+1\neq0
\end{array}
$
\\ \hline IViib(1) & $7$&$
\begin{array}{c}
T'_2\neq0,~A_5(t,r)=0,~\frac{T_2}{\sqrt{T_1}}(\frac{T'_2}{2T_2\sqrt{T_1}})'+1=0,\\
(\frac{T'_2}{2T_2\sqrt{T_1}})'\neq0
\end{array}
$
\\ \hline IViib2(*) & $4$&$
\begin{array}{c}
T'_2\neq0,~A_5(t,r)=0,~\frac{T_2}{\sqrt{T_1}}(\frac{T'_2}{2T_2\sqrt{T_1}})'+1=0,\\
(\frac{T'_2}{2T_2\sqrt{T_1}})'=0,~(\frac{T'_0}{T'_2})'\neq0
\end{array}
$
\\ \hline IViib2**(+) & $10$&$
\begin{array}{c}
T'_2\neq0,~A_5(t,r)=0,~\frac{T_2}{\sqrt{T_1}}(\frac{T'_2}{2T_2\sqrt{T_1}})'+1=0,\\
(\frac{T'_2}{2T_2\sqrt{T_1}})'=0,~\frac{T'_0}{T'_2}=\delta,~\delta>0
\end{array}
$
\\ \hline IViib2**(++) & $10$&$
\begin{array}{c}
T'_2\neq0,~A_5(t,r)=0,~\frac{T_2}{\sqrt{T_1}}(\frac{T'_2}{2T_2\sqrt{T_1}})'+1=0,\\
(\frac{T'_2}{2T_2\sqrt{T_1}})'=0,~\frac{T'_0}{T'_2}=\delta,~\delta=0
\end{array}
$
\\ \hline IViib2**(+++) & $10$&$
\begin{array}{c}
T'_2\neq0,~A_5(t,r)=0,~\frac{T_2}{\sqrt{T_1}}(\frac{T'_2}{2T_2\sqrt{T_1}})'+1=0,\\
(\frac{T'_2}{2T_2\sqrt{T_1}})'=0,~\frac{T'_0}{T'_2}=\delta,~\delta<0
\end{array}
$
\\ \hline IV(iii) & IVib(1)&$
\begin{array}{c}
T_2=constant,~A_5=0
\end{array}
$
\\ \hline
\end{tabular}
\end{center}
When we take one component non-zero (Case I), we obtain infinite
dimensional MCs for the degenerate case and one MC for the
non-degenerate case. For two components non-zero (Case II), the
degenerate case gives infinite dimensional Lie algebra while the
non-degenerate case yields maximum three MCs. The case III has four
different possibilities. For the first possibility, we obtain three
MCs and the second possibility gives four independent MCs. It is
mentioned here that finite results have been obtained for the
non-degenerate cases. The cases III(iii) and III(iv) are interesting
one where we obtain finite MCs even for the degenerate case. These
are five independent MCs. It is worth noticing that this (Case III)
generalizes the degenerate case of the static spherically symmetric
spcetimes [15]. We know that every KV is an MC, but the converse is
not always true. The last case (Case IV) generalizes the
non-degenerate case of the static spherically symmetric spacetimes
[15] for the case $k=1$. Here we take all the components of the
vector $\xi^a$ non-zero and solve the MC equations generally. We
obtain either four, five, six, seven or ten independent MCs in which
four are the usual KVs  and rest are the proper MCs.

Finally, it is remarked that RCs obtained by Akbar [24] are similar
to MCs. However, the constraint equations are different. If we solve
these constraint equations, we may have a family of spacetimes. It
would be interesting to look for solutions from these constraint
equations.

\renewcommand{\theequation}{A\arabic{equation}}
\setcounter{equation}{0}
\section*{Appendix A}

The surviving components of the Ricci tensor are
\begin{eqnarray}
R_{00}&=&(\frac{\nu''}{2}-\frac{\nu'\mu'}{4}
+\frac{\nu'^2}{4}+\frac{\nu'}{r})e^{\nu-\mu},\\
R_{11}&=&-\frac{\nu''}{2}+\frac{\nu'\mu'}{4}
+\frac{\nu'^2}{4}+\frac{\mu'}{r},\\
R_{22}&=&(\frac{r\mu'}{2}-\frac{r\nu'}{2}-1)e^{-\mu}+k,\\
R_{33}&=&f^2T_{22}.
\end{eqnarray}
The Ricci scalar is
\begin{eqnarray}
R=(\nu''-\frac{\nu'\mu'}{2}+\frac{\nu'^2}{2}-\frac{2\mu'}{r}
+\frac{2\nu'}{r}+\frac{2}{r^2})e^{-\mu}-\frac{2k}{r^2}.
\end{eqnarray}
The non-vanishing components of the energy-momentum tensor turn out
to be
\begin{eqnarray}
T_{00}&=&\frac{1}{r}[(\mu'-\frac{1}{r})e^{-\mu}+\frac{k}{r}]e^\nu,\\
T_{11}&=&\frac{1}{r}(\nu'-\frac{k}{r}+\frac{e^{\mu}}{r}),\\
T_{22}&=&\frac{r}{2}(\nu'-\mu'+r\nu''-\frac{r\nu'\mu'}{2}
+\frac{r\nu'^2}{2})e^{-\mu},\\
T_{33}&=&f^2T_{22}.
\end{eqnarray}

\newpage

\begin{description}
\item {\bf Acknowledgments}
\end{description}

I would like to thank Punjab University for the traveling grant to
visit Osaka City University, Japan and the host Institute for
providing local hospitality, where a part of this work was
completed.

\vspace{2cm}

{\bf \large References}

\begin{description}

\item{[1]} Katzin G.H., Levine J. and Davis W. R.: J. Math. Phys. {\bf
10}(1969)617.

\item{[2]} Oliver, D.R. and Davis, W.R.: Gen. Rel. Grav. {\bf
8}(1977)905; {\it Ann. Inst. Henri Poincare} {\bf 30}(1979)339.

\item{[3]} Collinson, C.D.: Gen. Rel. Grav. {\bf 1}(1970)137.

\item{[4]} Tsamparlis, M. and Mason, D.P.: J. Math. Phys. {\bf 31}(1990)1707.

\item{[5]} Davis, W.R., Green, L.H. and Norris, L.K.: Nuovo Cimento {\bf B34}(1976)256.

\item{[6]} Green, L.H, Norris, L.K. and Oliver Jr, D.R. and Davis, W.R.:
Gen. Rel. Grav. {\bf 8}(1977)731.

\item{[7]} Nunez, L., Percoco, U. and Villalba, V.H.: J. Math. Phys. {\bf 31}(1990)137.

\item{[8]} Melfo, Nunez, L.A., Percoco, U. and Villalba, V.H.:
J. Math. Phys. {\bf 33}(1992)2258.

\item{[9]} Bokhari, A.H. and Qadir, A.: J. Math. Phys. {\bf 34}(1993)3543;\\
Amir, M.J., Bokhari, A.H. and Qadir, A.: J. Math. Phys. {\bf 35}(1994)3005;\\
Bin Farid, T., Qadir, A. and Ziad, M.: J. Math. Phys. {\bf 36}(1995)5812.\\
Ziad, M.: Gen. Rel. Grav. {\bf 35}(2003)915.\\
Qadir, Asghar, Saifullah, K. and Ziad, M.: Gen. Rel. Grav. {\bf
35}(2003)1927.

\item{[10]} Yavuz, \.I and Camc{\i}, U, Gen. Rel. Grav.: {\bf 28}(1996)691.

\item{[11]} Carot, J. Nunez, L. and Percoco U. Gen. Rel. Grav. {\bf 29}(1997)1223.

\item{[12]} Tsamparlis, M., and Apostolopoulos, P.S.: Gen. Rel. Grav. {\bf
36}(2004)47.

\item{[13]} Carot, J., da Costa, J. and Vaz, E.G.L.R.: J. Math.
Phys. {\bf 35}(1994)4832.

\item{[14]} Hall, G.S., Roy, I. and Vaz, L.R.: Gen. Rel and Grav.
{\bf 28}(1996)299.

\item{[15]} Sharif, M.: Nuovo Cimento {\bf B116}(2001)673;
Astrophys. Space Sci. {\bf 278}(2001)447; J. Math. Phys. (2003);\\
Sharif, M. and Sehar Aziz: Gen Rel. and Grav. {\bf 35}(2003);\\
Sharif, M.: J. Math. Phys. {\bf 45}(2004)1518; ibid {\bf
45}(2004)1532; ibid {\bf 45}(2004)4193.

\item{[16]} Camc{\i}, U. and Sharif, M.: Gen Rel. and Grav. {\bf
35}(2003)97.

\item{[17]} Camc{\i}, U. and Sharif, M.: Class. Quant. Grav. {\bf
20}(2003)2169.

\item{[18]} Camci, U. and Barnes, A.: Class. Quant. Grav. {\bf 19}(2002)393.

\item{[19]} Camci, U. and Sharif, M.: Gen. Rel. Grav. {\bf 35}(2003)97.

\item{[20]} MacCallum, M.A.H., {\it In General Relativity:
An Einstein Centenary Survey},
eds. Hawking, S. and Israel, W.(Cambridge Univ. Press, 1979)533.

\item{[21]} Baofa, H.: Int. J. Theor. Phys. {\bf 30}(1991)1121.

\item{[22]} Lorenz, D. J.: Phys. A: Math. Gen. {\bf 15}(1982)2809.

\item{[23]} Kramer, D., Stephani, H., MacCallum, M.A.H. and
Hearlt, E.: {\it Exact Solutions of Einstein's Field Equations}
(Cambridge University Press, 2003).

\item{[24]} Akbar, M. and Rong-Gen, CAI: Commun. Theor. Phys. {\bf
45}(2006)95.\\
Akbar, M.: arXiv:gr-qc/0604030.

\end{description}

\end{document}